\documentclass[english]{emulateapj}
\usepackage[T1]{fontenc}
\usepackage[latin9]{inputenc}
\usepackage{amstext}
\usepackage{graphicx}
\usepackage{amsmath}
\usepackage{amssymb}
\usepackage{longtable}
\makeatletter

\providecommand{\tabularnewline}{\\}

\makeatother

\usepackage{babel}
\begin{document}

\title{Supernovae from direct collisions of white dwarfs \\
and the role of helium shell ignition }

\author{Oded Papish \& Hagai Binyamin Perets}

\affil{Technion - Israel Institute of Technology, physics department, Haifa,
Israel 32000}
\begin{abstract}
Models for supernovae (SNe) arising from thermonuclear explosions
of white dwarfs (WDs) have been extensively studied over the last
few decades, mostly focusing on the single degenerate (accretion of
material of a WD) and double degenerate (WD-WD merger) scenarios.
In recent years it was suggested that WD-WD direct collisions provide
an additional channel for such explosions. Here we extend the studies
of such explosions, and explore the role of Helium-shells in affecting
the thermonuclear explosions. We study both the impact of low-mass
helium ($\sim0.01$ M$_{\odot})$ shells, as well as high mass shells
($\ge0.1$ M$_{\odot}$). We find that detonation of the massive helium
layers precede the detonation of the WD Carbon-Oxygen (CO) bulk during
the collision and can change the explosive evolution and outcomes
for the cases of high mass He-shells. In particular, the He-shell
detonation propagates on the WD surface and inefficiently burns material
prior to the CO detonation that later follows in the central parts
of the WD. Such evolution leads to larger production of intermediate
elements, producing larger yields of $^{44}{\rm Ti}$ and $^{48}{\rm Cr}$
relative to the pure CO-CO WD collisions. Collisions of WDs with a
low-mass He-shell do not give rise to helium detonation, but helium
burning does precede the CO bulk detonation. Such collisions produce
a high velocity, low-mass of ejected burned material enriched with
intermediate elements, with smaller changes to the overall explosion
outcomes. The various effects arising from the contribution of low/high
mass He layers change the kinematics and the morphological structure
of collision-induced SNe and may thereby provide unique observational
signatures for such SNe, and play a role in the chemical enrichment
of galaxies and the production of intermediate elements and positrons
from their longer-term decay. 
\end{abstract}

\section{Introduction}

Thermonuclear explosions of white dwarfs (WDs) have been extensively
studied over the last decades, and are thought to be the progenitors
of various types of supernovae (SNe), most notably type Ia SNe, but
possibly also calcium-rich faint type Ib SNe \citep{per+10}. Models
for thermonuclear detonations are traditionally divided between double
degenerate (DD) models and single degenerate models (SD). The former
focus on cases of mergers of two WDs, and the latter on accretion
of material from a companion star onto a WD (See \citealp{mao+14}
for a review). In recent years it was suggested that the DD scenario
can be extended to WD-WD \textit{collisions} and not only mergers,
but the former were thought to be extremely rare, and occur only in
dense stellar clusters. For this reason they attracted relatively
little attention compared with other WD explosion progenitors. Such
collisions, however, are likely to be observable as type Ia SNe \citep{2009ApJ...705L.128R}
and possibly non-standard SNe \citep{2009MNRAS.399L.156R,2010ApJ...724..111R}.
Recently, it was shown that some triple systems may dynamically evolve
through a quasi-secular process, reminiscent of Kozai-Lidov oscillations
\citep{koz62,lid62}, but where significant peri-center changes can
occur on a single orbit time-scale \citet{2012ApJ...757...27A}, leading
to extremely close peri-center approaches. In particular, it was suggested
that such evolution in triples hosting an inner WD-WD binary could
lead to physical collisions and the production of type Ia SNe \citep{2011ApJ...741...82T,kat+12,2013ApJ...778L..37K}.
Though such systems are not likely to be frequent (e.g. \citet{2013MNRAS.430.2262H};
see also section \ref{sub:rates} in the following), they may be much
more frequent than WD-WD collisions in dense clusters, making such
scenarios more likely. 

While previous studies mostly focused on collisions of carbon-oxygen
(CO), the potential role of a He component, and in particular He-shells
on CO-WDs attracted only little attention. \citet{2014ApJ...785..124K}
calculated the conditions for a detonation in a tiny helium shell
around two $1M_{\odot}$ WDs and concluded that detonation cannot
occur during the collision. Realistic CO WDs typically hold outer
Helium layers which can range in mass between $10^{-3}$ ${\rm M_{\odot}}$
for more massive WDs up to 0.024 ${\rm M_{\odot}}$ for the lowest
mass CO WDs \citep{law+06}. In the case of an interacting binary
where a WD grows through accretion, the accreting WD might have a
significant outer Helium layer of up to $0.1$ ${\rm M_{\odot}}$
or more on low mass WD (e.g. \citealt{bil+07}); such a helium layer
is thought to play a major role in the SD sub-Chandrasekhar and .Ia
models for thermonuclear explosions \citep{woo+86,liv+91,bil+07,she+09,per+10,wal+11};
the helium layer may ignite and later possibly induce the detonation
of the CO-WD bulk through converging shock \citep{2014ApJ...785...61S}.
Recently, it was suggested to play a similar role in WD-WD mergers
\citet{2013ApJ...770L...8P}. Here we study the outcomes of WD-WD
collisions and explore the possibility that a helium shell can play
an important role in such collisions. As we discuss in the following,
a He-shell can produce non-negligible effects on the collisions outcome,
changing the composition, energetics and velocity distributions of
the resulting SN; the amplitude of these effects depend on the mass
of the He layer considered, with more significant effects arising
from larger He layers. 

We begin by describing our 2D simulations and their initial conditions
(section \ref{sec:Simulations}), and present the main results in
section \ref{sec:Results}. We then discuss our results (section \ref{sec:Discussion}),
where we also discuss the rates of such collisions and point out their
main implications and their observational signatures.

\section{Simulations}
\label{sec:Simulations}

\subsection{WD Models}
\label{sub:WD-Models}

The WD models are calculated using the MESA stellar evolution code
\citep{2011ApJS..192....3P}. The initial WD models were taken from
\citet{2013ApJ...777..136W} and then changed using MESA for the different
compositions. The models with a small helium shell were built by relaxing
the chemical composition of the non helium elements to be 50\% carbon
and 50\% oxygen. For the CO models we removed the helium shell up
to a total of $10^{-3}M_{\odot}$. After that we relaxed the WDs composition
to be 50\% carbon and 50\% oxygen. The models with a large helium
shell were built by accretion at a rate of $10^{-8}M_{\odot}{\rm yr}^{-1}$
of helium to the initial models up to a total $0.1M_{\odot}-0.2M_{\odot}$
of helium on the WDs. In the following we consider the whole
range of low-mass to high mass helium shells, but we should note that
the latter cases of massive shells are likely to be rare. Nevertheless,
even massive helium shells may exist around WDs in binaries with a
helium donor star (e.g. AM-CVn systems), and the possibility of helium
shell detonation in such cases has been widely explored \citep[e.g.][and references therein]{woo+86,liv+91,wal+11}.
A collision of a CO WD with such a WD with a massive helium shell
could therefore occur either randomly in dense clusters, or in high
multiplicity systems (>2 component) in which one of the components
is such a helium accreting WD in a close binary, and another outer
CO WD companion exists (e.g. \citealp{Krz+72} find a triple system
with an inner AM CVn). The outer WD do not need to be close, as even
collisions with very far companions could occur in the field \citep{mic+15}.

\subsection{Hydrodynamical Simulations}
\label{sub:Hydrodynamical-Simulations}

We carried 17 different 2D simulations of WD-WD head-on collisions
using the FLASH v4.2 code \citep{2000ApJS..131..273F}. The widely
used FLASH code is a publicly available code for supersonic flow suitable
for astrophysical applications. The simulations were done using the
unsplit PPM solver of FLASH in 2D axisymmetric cylindrical coordinates
on a grid of size $3.2\times6.4\;\left[10^{10}\,{\rm cm}\right]$
using adaptive mesh refinement (AMR). To prevent the production of
unreal early detonation that may arise from numerical resolution we
applied a limiter approach following \citet{2013ApJ...778L..37K}.
We made multiple simulations with increased resolution until convergence
was reached in the nuclear burning. We found a resolution of $5-10\,{\rm km}$
to be sufficient for convergence to up to $10\%$. Gravity was included
as a multipole expansion of up to multipole $l=10$ using the new
FLASH multipole solver. The equation of state used in the simulations
is the Helmholtz EOS \citep{2000ApJS..126..501T}. This EOS includes
contributions form partial degenerate electrons and positrons, radiation,
and non degenerate ions. The nuclear network was FLASH's 19 isotopes
alpha chain network. This network can capture well the energy generated
during the nuclear burning \citep{2000ApJS..129..377T}. As initial
conditions we take the two WDs to be at contact with free fall velocities.
The various simulations conducted are summarized in table \ref{tab:Collisions-models}.

\begin{table}[t]
\begin{tabular}{lllll}
\hline 
\hline 

Model & Primary &  & Secondary & \tabularnewline
 &  Mass {[}$M_{\odot}${]} & $\rho_{{\rm c}}$ {[}g{]} &  Mass {[}$M_{\odot}${]} & $\rho_{{\rm c}}$ {[}g{]}\tabularnewline
\hline 
\hline 
1A & 0.6 CO  & $3.4\times10^{6}$ & 0.6 CO  & $3.4\times10^{6}$\tabularnewline
1B & 0.6 CO  & $3.4\times10^{6}$ & 0.6 CO + 0.01 He  & $3.5\times10^{6}$\tabularnewline
1C & 0.6 CO + 0.01 He & $3.5\times10^{6}$ & 0.6 CO +0.01 He  & $3.5\times10^{6}$\tabularnewline
1D & 0.6 CO + 0.024 He  & $4.0\times10^{6}$ & 0.6 CO +0.024 He  & $4.0\times10^{6}$\tabularnewline
1E & 0.6 CO  & $3.4\times10^{6}$ & 0.6 CO + 0.1 He  & $5.7\times10^{6}$\tabularnewline
1F & 0.6 CO + 0.1 He & $5.7\times10^{6}$ & 0.6 CO + 0.1 He & $5.7\times10^{6}$\tabularnewline
1G & 0.6 CO + 0.2 He & $1.1\times10^{7}$ & 0.6 CO + 0.2 He & $1.1\times10^{7}$\tabularnewline
\hline 
2A & 0.7 CO  & $5.9\times10^{6}$ & 0.7 CO  & $5.9\times10^{6}$\tabularnewline
2B & 0.7 CO  & $5.9\times10^{6}$ & 0.7 CO + 0.01 He  & $6.1\times10^{6}$\tabularnewline
2C & 0.7 CO + 0.01 He  & $6.1\times10^{6}$ & 0.7 CO +0.01 He  & $6.1\times10^{6}$\tabularnewline
2D & 0.7 CO + 0.024 He  & $6.8\times10^{6}$ & 0.7 CO +0.024 He  & $6.8\times10^{6}$\tabularnewline
2E & 0.7 CO  & $5.9\times10^{6}$ & 0.7 CO + 0.1 He  & $9.9\times10^{6}$\tabularnewline
\hline 
3 & 0.6 CO  & $3.4\times10^{6}$ & 0.7 CO  & $5.9\times10^{6}$\tabularnewline
4 & 0.7 CO  & $5.9\times10^{6}$ & 0.8 CO  &  $ 1.0 \times 10^{7}$\tabularnewline
\hline 
5A & 0.8 CO  &  $ 1.0 \times 10^{7}$ & 0.6 CO  & $3.4\times10^{6}$\tabularnewline
5B & 0.8 CO  &  $ 1.0 \times 10^{7}$ & 0.6 CO + 0.01 He  & $3.5\times10^{6}$\tabularnewline
5C & 0.8 CO  &  $ 1.0 \times 10^{7}$ & 0.6 CO + 0.1 He  & $5.7\times10^{6}$\tabularnewline
\hline 
 &  & \tabularnewline
\end{tabular}

\caption{\label{tab:Collisions-models}Collision models.}
\end{table}

\section{Results}
\label{sec:Results}

\subsection{Collisions of pure CO-WDs and comparison with previous collision
simulations}

We run several pure CO-WDs collisions in order to check the consistency
of our simulations with previous works, and then use them as a benchmark
for comparison with the novel collision simulations in which WDs with
He-layers were considered. 

Previous collision simulations include several works by different
groups \citep{2009ApJ...705L.128R,2009MNRAS.399L.156R,2010ApJ...724..111R,2010MNRAS.406.2749L,2012ApJ...759...39H,2013ApJ...778L..37K,2013MNRAS.436.3413G}.
In these simulations only pure CO-WD collisions were simulated (or
CO-WD - He-WD collisions \citealt{2009ApJ...705L.128R}). Early studies
of WD-WD collisions showed large discrepancies in the $^{56}{\rm Ni}$
yields produced, likely resulting from insufficient resolution \citep{2013ApJ...778L..37K}
and the use of too-large time steps in coupling the nuclear burning
to the hydrodynamics \citep{2012ApJ...759...39H}. \citet{2013ApJ...778L..37K}
devised the limiter approach to tackle such potential difficulties;
recent high resolutions (e.g., \citealt{2013MNRAS.436.3413G}) show
a convergence to \citet{2013ApJ...778L..37K} results, supporting
the validity of their limiter approach. Our main comparison is therefore
with the results by \citet{2013ApJ...778L..37K}, who similarly used
the FLASH code, and included the limiter approach. One difference
in our simulations is the use of the unsplit hydro solver in FLASH
while \citealt{2013ApJ...778L..37K} uses the split solver. 

Our pure CO-WDs collisions result in CO detonations in both WDs. The
total energy $E_{k}$ and abundances of helium, intermediate elements,
and nickel are presented in Table \ref{table2}. All the results are
shown after the nuclear reactions in the simulations ended. $E_{k}$
includes kinetic, thermal, and gravitational energy. The amount of
nickel and total energy we get is similar to the results of \citet{2013ApJ...778L..37K}.
This is expected as in both works the same code is used with the same
limiter on the nuclear reaction rate. We find our results to be similar
up to $\sim10\%$ in the pure CO cases. These small differences should
exist as a result of the slightly different initial preparation of
the progenitors and the resolutions used in the two studies. We therefore
confirm previous studied of such collisions and find them to be consistent
with the results we obtain

\subsection{Collisions between CO-WDs with a low-mass, non-detonating helium
layer}

\begin{figure}[b]
\begin{centering}
\includegraphics[scale=0.6]{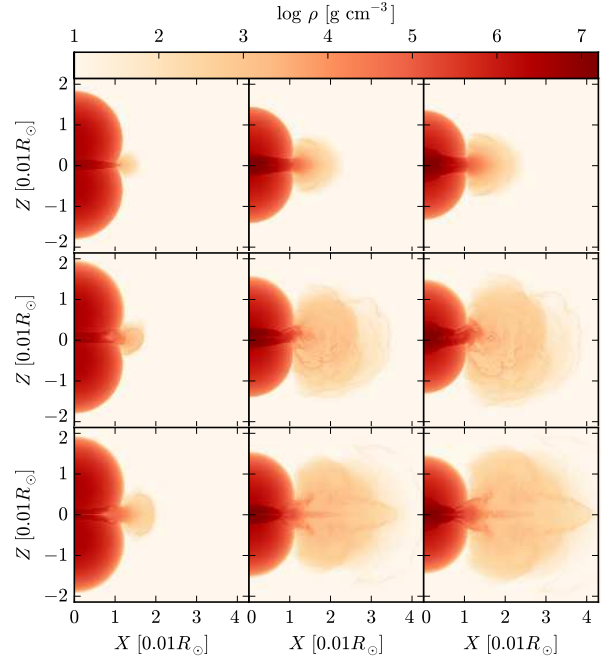}
\par\end{centering}

\caption{\label{fig:dens-06}Density before,during, and after CO detonation
for two $0.6M_{\odot}$ WDs. Upper panels: pure CO WDs. Middle panels:
the top WD with $0.01M_{\odot}$ of helium while the bottom WD is
a pure CO. Bottom panels: both WDs with a layer of $0.01M_{\odot}$
of helium. The detonation in the CO is similar in the three cases.}
\end{figure}

Collisions of WDs with low-mass helium layers (up to $M_{{\rm He}}<0.024M_{\odot}$;
see Fig. \ref{fig:dens-06}) do not give rise to a detonation in the
helium layer. Such collisions are characterized by three phases: a)
The helium compression phase. A short time after the collision
the helium layer is compressed to densities of $\rho_{{\rm He}}\approx10^{5}\,{\rm g}{\rm cm^{-3}}$
and temperatures of $T_{{\rm He}}\approx10^{9}\,{\rm K}$. Under these
conditions the helium stars burning, but no detonation is triggered.
The compressed helium is ejected in the plane perpendicular to the
collision direction with velocities of up to $20,000\,{\rm km}\,{\rm s^{-1}}$
as it burns to intermediate elements. b) The CO detonation
phase. This phase is very similar to the pure CO-WDs collisions.
The compression and burning of the helium layer has almost no effect
on the CO core. The detonation in the CO material is triggered in
a very similar way to pure CO-CO collisions \citep{2013ApJ...778L..37K}
and can be seen in Fig. \ref{fig:dens-06}. c) Outer helium
layer ejection. The detonation wave in the CO material reaches the
outer part of the helium layer and eject the partially burned helium
layer. At the end of the explosion $\sim1/2$ of the initial helium
mass survives in all cases and is ejected after the shock reaches
the outer part of the WD. Intermediate mass elements are also ejected
in this way. The $^{56}{\rm Ni}$ yields in all cases are comparable
to the $^{56}{\rm Ni}$ yield in pure CO-CO collisions. The total
mass of the intermediate elements is summarized in table \ref{table2}. 
We caution that exact outcomes of
these simulations depend on the nucleosynthetic network used. Here
we use the 19 isotopes alpha chain network; in section \ref{sub:Nuclear-reactions-post-processin}
we show that although the simplified smaller network well captures
the overall evolution, energetics and nucleosynthetic production,
the result obtained differ quantitatively (and non-negligibly) from
those obtained using a much larger network. In particular, the latter
show an increase of the nickel production and suppression of the IME
production, compared with the results of the smaller network.

\begin{table}[h]
\begin{tabular}{lccccccc}
\hline 
\hline 
 &  &  &  &  &  &  & Helium\tabularnewline

Model & $^{4}{\rm He}$ & $^{40}{\rm Ca}$ & $^{44}{\rm Ti}$ & $^{48}{\rm Cr}$ & $^{56}{\rm Ni}$ & $E_{\rm {exp}}$ & Detonation \tabularnewline
\hline 
1A & 1(-3) & 4(-2) & 9(-5) & 4(-4) & 0.31 & 1.3  & - \tabularnewline
1B & 5(-3) & 4(-2) & 1.5(-3) & 1(-3) & 0.27 & 1.3 & No\tabularnewline
1C & 1(-2) & 5(-2) & 3(-3) & 2(-3) & 0.29 & 1.3 & No\tabularnewline
1D & 3(-2) & 4(-2) & 5(-3) & 3(-3) & 0.38 & 1.2 & Yes\tabularnewline
1E & 4(-2) & 4(-2) & 1.3(-2) & 1(-2) & 0.4 & 1.4 & Yes\tabularnewline
1F & 1(-1) & 4(-2) & 2(-2) & 2(-2) & 0.8 & 1.8 & Yes\tabularnewline
1G & 1(-1) & 6(-2) & 7(-3) & 9(-3) & 0.3 & 2.1 & Yes\tabularnewline
\hline 
2A & 3(-3) & 5(-2) & 1(-4) & 6(-4) & 0.52 & 1.5 & -\tabularnewline
2B & 6(-3) & 4(-2) & 8(-4) & 9(-4) & 0.65 & 1.7 & No\tabularnewline
2C & 8(-3) & 5(-2) & 1.5(-3) & 1(-3) & 0.54 & 1.6 & No\tabularnewline
2D & 3(-2) & 5(-2) & 3(-3) & 3(-3) & 0.61 & 1.6 & Yes\tabularnewline
2E & 5(-2) & 5(-2) & 1.4(-2) & 1(-2) & 0.67 & 1.6 & Yes\tabularnewline
\hline 
3 & 2(-3) & 4(-2) & 1(-4) & 5(-4) & 0.42 & 1.4 & -\tabularnewline
4 & 1(-3) & 4(-2) & 1.6(-4) & 6(-4) & 0.49 & 1.6 & -\tabularnewline
\hline 
5A & 2(-3) & 3(-2) & 4(-5) & 3(-4) & 0.32 & 1.4 & -\tabularnewline
5B & 7(-3) & 4(-2) & 1(-3) & 6(-4) & 0.27 & 1.4 & No\tabularnewline
5C & 5(-2) & 6(-2) & 1.5(-2) & 1(-2) & 0.4 & 1.7 & Yes\tabularnewline
\end{tabular}
\label{table2}
\protect\caption{Abundances at the end of the simulations in values of $M_{\odot}$. Energy is in $10^{51} ~\rm{erg}$.}
\end{table}

\begin{figure}[h]
\begin{centering}
\includegraphics[scale=0.65]{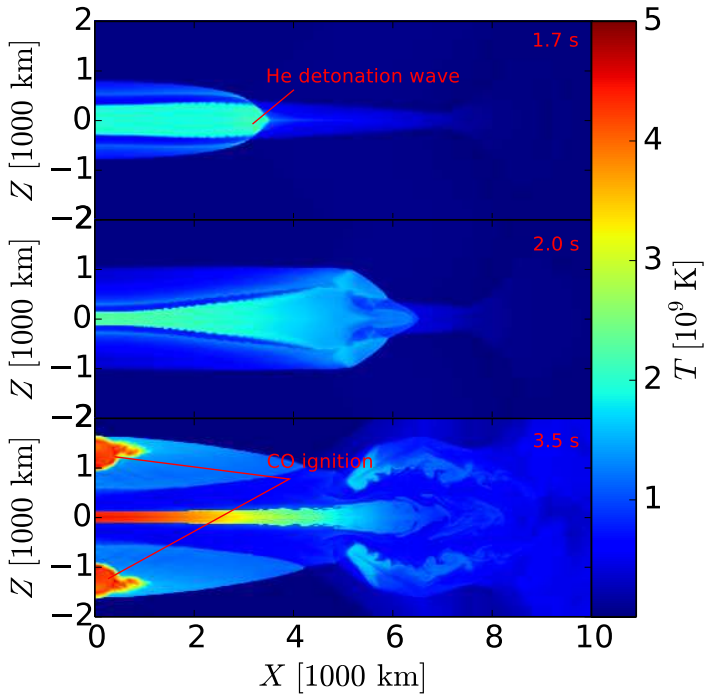}
\par\end{centering}

\caption{\label{fig:CO-Ignition}Ignition in the He layer and C/O core of two
WDs with composition of $0.6M_{\odot}$ C/O and $0.024M_{\odot}$
of He. The helium layer is detonated first resulting in a fast ejecta
in the horizontal plan. The helium detonation front is not able to
propagate along the curved helium shell. The helium ignition has no
real effect on the CO ignition process. }
\end{figure}

\begin{figure}[h]
\begin{centering}
\includegraphics[scale=0.6]{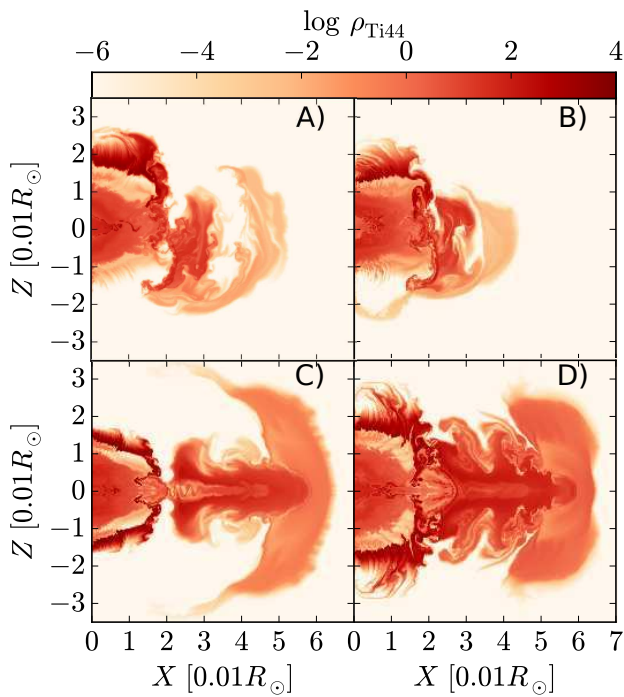}
\par\end{centering}

\caption{\label{fig:ti44-06}Density of $^{44}{\rm Ti}$ after the shock reaches
the outer parts of the WDs. A) Top WD is a $0.6M_{\odot}$CO with
$0.01M_{\odot}$ of helium; bottom WD is a pure $0.6M_{\odot}$ CO
WD. B) Top WD is a $0.7M_{\odot}$CO with $0.01M_{\odot}$ of helium;
bottom WD is a pure $0.7M_{\odot}$ CO WD . C) Both WDs are $0.6M_{\odot}$CO
with $0.01M_{\odot}$ of helium. D) Both WDs are $0.6M_{\odot}$CO
with $0.024M_{\odot}$ of helium. Most of the ejected $^{44}{\rm Ti}$
is from the outer parts of the WDs from partial burning of the helium
layer. A small plump is ejected during the collision and can be seen
in front of the WDs. }
\end{figure}

\subsection{Induced helium detonation in collisions between CO-WD with a large-mass
helium layer}

\begin{figure*}
\begin{centering}
\includegraphics[scale=0.75]{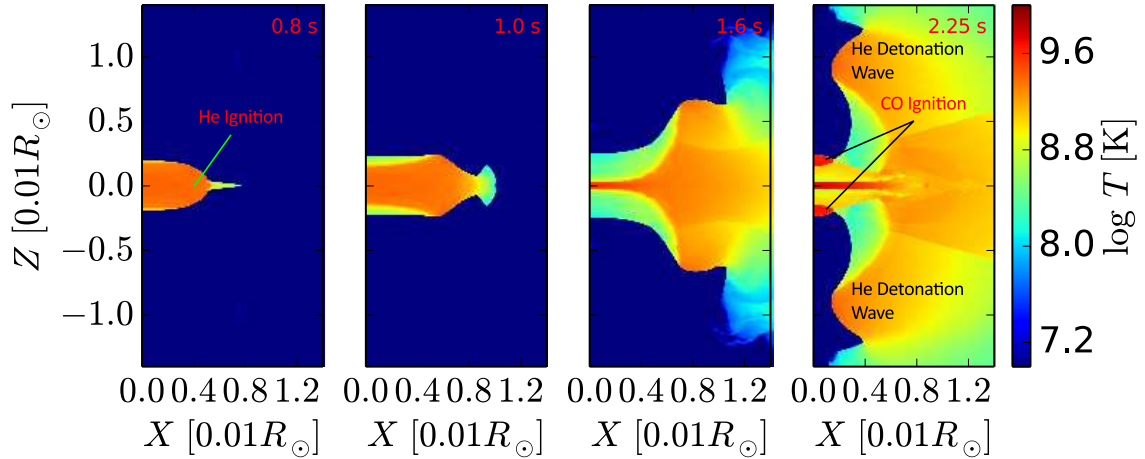}
\par\end{centering}

\caption{\label{fig:ignition-01}Ignition in the Helium shell and the CO core
of two WDs with composition of $0.6M_{\odot}$ CO and $0.1M_{\odot}$
of He. The helium layer detonates first and propagate along the outer
shell. The CO core is then ignited as a result of the collision's
shock. }
\end{figure*}

In collisions of WDs with a massive ($M_{{\rm He}}\ge0.024M_{\odot}$)
helium layers, the helium detonates during its compression phase.
For the $M_{{\rm He}}=0.024M_{\odot}$ case the helium detonation
is not able to propagate through the curved helium shell (Fig. \ref{fig:CO-Ignition})
and has almost no influence on the overall collision outcomes, i.e.
the detonation in the CO core is triggered by the collision in a similar
way to pure CO-WDs collisions. For more massive shells ($M_{{\rm He}}=0.1M_{\odot}$; 
models 1F,2E) 
the helium detonation is able to propagate through the shell before
the CO is detonated (see Fig. \ref{fig:ignition-01}). In this case
we find that the helium shell burning results in the production of
non-negligible amounts of intermediate mass elements and nickel. The
velocity of the helium-burned material is lower than the velocity
of the material ejected following the CO core detonation which gradually
outruns the material ejected through the earlier He-detonation in
the shell.  For even more massive helium shells (($M_{{\rm He}}=0.2M_{\odot}$;
 model 1G) 
the converging shock caused by the helium detonation is able to detonate
the CO. In this case the induced CO detonation is obtained earlier
than the corresponding CO detonation in the pure CO-CO WD collisions
and both the nickel abundance ($M_{{\rm Ni}^{56}}\simeq0.3M_{\odot})$
and the abundance of intermediate elements are smaller than in the
pure CO collisions. 

\subsection{Nuclear reactions post-processing}
\label{sub:Nuclear-reactions-post-processin}
In order to provide a better understanding of the nucleosynthetic
yields from WD-WD collisions we go beyond previous models using a
small nuclear reaction netweorks networks. We extend these by including
a large 160 post-processed nuclear reaction network for two of our
simulations, a pure CO-CO collision (1A) and a CO+He - CO+He (low-mass
helium shell) collision (1C). The post-processing was done by including
30,000 trace particles to the simulations and record their thermodynamical
conditions throughout the simulations. We post-processed the results
from the trace particles using the burn unit of the MESA code (version
7624) \citep{2015ApJS..220...15P}. MESA uses the JINA reaclib version
V2.0 2013-04-02 \citep{2010ApJS..189..240C} database for nuclear
reaction rates. The results are summarized in table \ref{post-processing}.

The results show production of a larger amount of $^{56}$Ni
and suppression of $^{44}$Ti with respect to the results obtained
using the smaller $\alpha$ network in \textsc{FLASH}
(19 . Similar differences due to the use of larger nucleosynthetic
network have been seen in other works \citep{2013MNRAS.436.3413G},
which arise due to the more efficient nuclear burning, producing larger
abundances of higher elements (e.g., $^{56}$Ni) and lower
amounts of intermediate elements (e.g., $^{44}$Ti).

\subsection{Influence of initial separation of the WDs}

The simulations in this study were initiated when the two
WDs are at contact, and the effect of tidal distortion of the WDs
prior to contact were not taken into account. To verify whether such
pre-impact effect play a role in the collision outcomes we ran an
additional simulation, identical to 1C, but with an initial separation
of $3\left(R_{{\rm WD}1}+R_{{\rm WD2}}\right)$. The results show
vary little compares with those obtained from `initial contact` simulations,
in particular there only a very small distortion is observed in the
WDs structure prior to impact. This is anticipated as the free fall
velocity is supersonic and the WDs do not have sufficient time to
be significantly distorted by much before the collision. In summary,
the energetics, overall evolution post impact and nucleosynthetic
yields show only very slight deviations from the `initial contact`
simulations.

\section{Discussion}
\label{sec:Discussion}

\subsection{Composition, morphology and kinematics}

\begin{figure}
\begin{centering}
\includegraphics[scale=0.4]{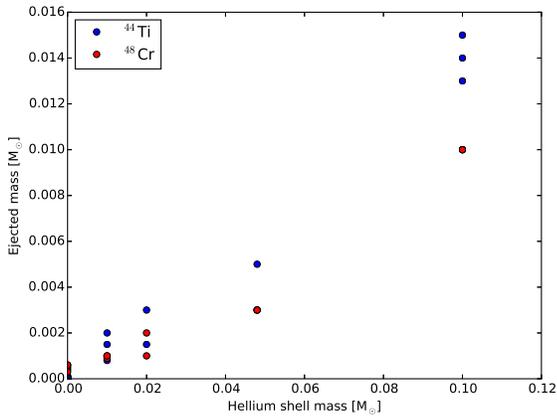}
\par\end{centering}

\caption{\label{fig:ti-cr-ejecta} Mass ejecta of $^{44}{\rm Ti}$ and $^{48}{\rm Cr}$
for the different models. The lower left points are for pure CO-WD
collisions. }
\end{figure}

In our simulations of WD-WD collisions including a He-layer we find
that the helium is burned mostly into intermediate elements resulting
in a much higher mass ejecta of these elements compared with pure
CO collisions. Only a small fraction of the helium shell burns into
$^{56}{\rm Ni}$, which has only a minor effect on the $^{56}{\rm Ni}$
yields. The deviation in the yield of intermediate elements compared
with their yield in pure CO collisions is found to be almost linearly
dependent on the He mass (Fig. \ref{fig:ti-cr-ejecta}). Using LS
fit we find the relation $M_{{\rm he}4}=0.001883+0.462991M_{{\rm shell}}$,
i.e., almost half of the helium is left unburned. For $^{44}{\rm Ti}$
and $^{48}{\rm Cr}$ we find $M_{{\rm ti44}}=\left(0.132099\pm0.006176\right)M_{{\rm shell}}$,
$M_{{\rm cr48}}=\left(0.094573\pm0.003914\right)M_{{\rm shell}}$.

For helium shells with mass $M_{{\rm shell}}<10^{-3}M_{\odot}$
these yields are significantly higher than those produced in collisions
of bare CO white dwarfs. For low mass shells the helium ejecta mostly
burns as a result of the CO detonation shock and the ejecta is relatively
spherically symmetric besides a small ejected plume of He/He-burned
material (see Fig. \ref{fig:ti44-06}). However, a bipolar structure
is formed when the two helium shells on the corresponding two WDs
have different masses (Fig \ref{fig:ti44-06}). In the high mass cases
most of the burning results from the helium detonation wave. A large
fraction of intermediate elements are then ejected in a cone of $\approx30^{\circ}$
to the collision plane. 

\begin{figure*}
\begin{centering}
\includegraphics[scale=0.75]{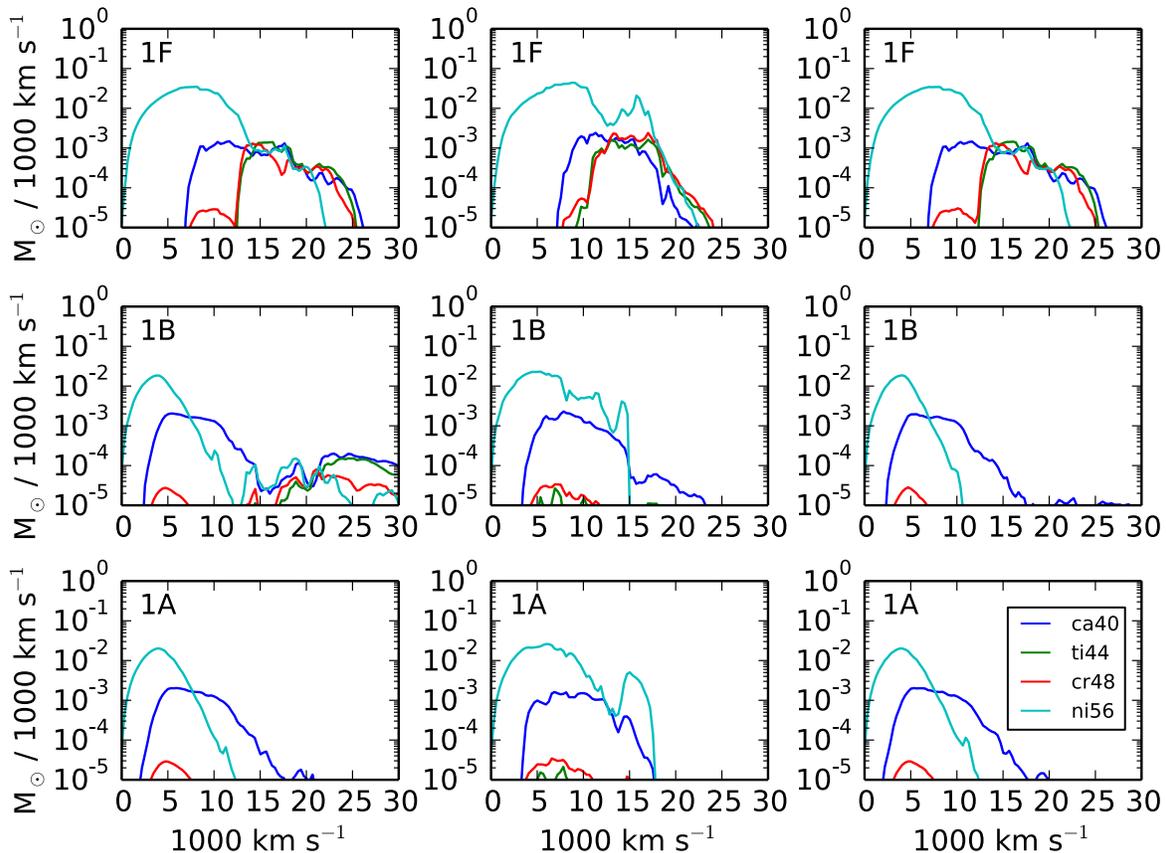}
\par\end{centering}

\caption{\label{fig:1E-v}Velocity profiles of $^{40}{\rm Ca}$, $^{44}{\rm Ti}$,
$^{48}{\rm Cr}$, $^{56}{\rm Ni}$ from three representing simulations.
The simulations are for the cases: 1A - no helium, 1B - one WD with
$0.01M_{\odot}$ of helium, 1F - two WDs with $0.1M_{\odot}$ of helium.
Left: lower region with $\cos\theta<-1/3$. Middle: middle region
with $1/3<\cos\theta<1/3$ . Right: upper region with $\cos\theta>1/3$.
The angle $\theta$ is the polar angle from the symmetry axis.}
\end{figure*}

The velocity distribution of the various components also differs between
the different type of collisions. In Fig. \ref{fig:1E-v} we present
the velocity profiles for cases 1A, 1B, 1F in the upper, middle, and
lower parts of the simulation domain. Collisions which include He-layers
show high velocity tails of ejecta of intermediate elements. In particular
the most He-rich collisions give rise to a non-negligible mass of
material ejected at higher velocities due to the He-detonation stage. 
Using a much larger nuclear network (see section \ref{sub:Nuclear-reactions-post-processin})
we find a reduction of IME production and an increase of nickel production,
showing that a large nuclear network is required for producing the
correct abundances (see also \citealp{Kushnir+15} for pointing out
these issues in a similar context).

\subsection{Comparison with other models of type Ia}

The overall energetics and behavior of the WD-collision models can
be relatively similar to other models for type Ia SNe. However, high
yields of intermediate elements do not show in models of the single
degenerate scenario. For example, model W7 of \citet{1984ApJ...286..644N}
results with $<10^{-4}M_{\odot}$of $^{44}{\rm Ti}$ and $<10^{-5}M_{\odot}$
of $^{48}{\rm Cr}$. As the amount of helium on a Chandrasekhar mass
WD is small we do not expect high yields of these elements in any
case except maybe for a failed Ia supernova (e.g., \citealt{2012ApJ...761L..23J,2013MNRAS.429.2287K};
this result might also is apply to the core-degenerate scenario \citealt{2003ApJ...594L..93L,2011MNRAS.417.1466K}). 

In the double degenerate scenario a helium shell can exist on the
WDs. \citet{2012ApJ...746...62R} simulated the merger of two WDs
with a helium layer; they found that a detonation occurs in the helium
layer when the primary mass was $1.06M_{\odot}$; however it did not
lead to carbon detonation as a result of the helium detonation. \citet{2013ApJ...770L...8P}
found that a violent merger with a $0.01M_{\odot}$helium shell produces
less than $2\times10^{-8}M_{\odot}$ of elements more massive than
calcium in the helium shell.

The relatively large abundances of intermediate elements such as $^{44}{\rm Ti}$
can provide a more unique signature for He-rich collisions, though
high abundances of these components can also be produced in the SD
sub-Chandrasekhar and .Ia explosion models \citep{woo+86,liv+91,bil+07,she+09,per+10,wal+11}).
Interestingly, recent observations \citep{tro+14} suggest that the
$^{44}{\rm Ti}$ abundances in Tycho SN remnant are comparable with
the abundances we find in the Helium rich collisions (even those with
low-mass He-layer). \citet{Lop+15} did not detect any $^{44}$Ti
but only derived upper limits of $2.4\times10^{-4}$ M$_{\odot}$.
Such $^{44}$Ti levels are significantly higher than expected
in typical SD scenarios (and much larger than the W7 model), but
are still consistent with the main models (1B,2B); {and are
lower than the sub-Chandrasekhar models (depending on the
assumed distance of the Tycho SNR; see figure 4 in \citealp{tro+14}).
If $^{44}$Ti rich SNe \emph{are} frequent, then their large production
of $^{44}{\rm Ti}$ could also be important for the 511 keV emission
in the Galaxy due to positron-electron annihilation following $^{44}{\rm Ti}$
decay (see \citealp{per14} for additional discussion of He-rich SNe
contribution).

\subsection{The rate of WD-WD collisions \label{sub:rates}}

The rate of WD-WD collisions in dense stellar systems is low, and
the overall rate was suggested to be dominated by possible WD-WD collisions
in quasi-secular triples. Estimating the rate of WD-WD collisions
is complicated and involves many uncertainties, both in the initial
fractions and orbital parameters of triple systems as well as the
effects of dynamical and stellar evolution. Such studies are beyond
the scope of this paper which focuses on the properties of the explosions
themselves. Nevertheless, one can constrain an upper limit on such
rates through simple arguments (see also \citealp{sok+14}).

Let us consider the most optimistic case. If we assume the binary
components in wide binaries are formed independently (uncorrelated)
then the mass-function of each stellar companion is independent. In
such a case, the fraction of WD companions to WDs among WD-binaries
is of the order of the overall WD fraction among single stars, namely
$\sim10\%$. If all of these were members of quasi-secular triple
systems, with isotropic distribution of mutual inclinations between
the inner and outer binaries of the triples, then $\sim5\%$ of these
inner WD-WD binaries will collide due to quasi-secular evolution \citep{kat+12},
with about half of the collisions having impact parameter sufficiently
small (smaller than half the WD radius), as to lead to a detonation
and a full thermonuclear explosion. In other words, $0.25\%$ of WD
systems will produce a thermonuclear explosion. Given that $\sim1\%$
of WD are required to explain the rates of type Ia SNe, then in the
most optimistic and highly simplistic case WD collisions in triples
can explain 25 \% of type Ia SNe. 

In practice, current available data suggest the rates are significantly
smaller. Among $0.9-1.5$ M$_{\odot}$ stars (constituting half of
the potential WD progenitors in the mass range $0.9-8$ M$_{\odot}$),
the triple and higher multiplicity fraction is only $13\%$ \citep{tok14,rid+15}.
Moreover, among these, only a small fraction$(<10\%$ ) of the systems
have period ratios residing in the quasi-secular regime ($P_{\text{out}}/P_{in}<\sim20)$.
Finally, there is reason to believe that the mutual inclinations of
at least the shorter period triples ($<100$ yrs), are relatively
small \citep{fek81}, rather than having an isotropic distribution.
Such low inclination triples will not contribute at all to the quasi-secular
regime. Taken together, one may conclude that WD progenitor systems
in lower mass range may give rise to $\sim100$ times \emph{smaller}
rates than the most optimistic estimated rate considered above. Even
allowing for a few times higher fractions due to some unaccounted-for
incompleteness, these observational data already suggest very low
rates. Moreover, some of the potential triple progenitors will induce
a merger/collision of the inner binary before its evolution into a
WD-WD binary. More massive stars are known to have higher multiplicity
fractions \citep{duc+13,san+13}, and the rates might therefore be
higher in the upper half of the mass range. However, such transition
in the multiplicity and orbital properties with mass is likely to
be continuous, and not produce orders of magnitude higher fractions
of quasi-secular triples than those seen for the lower mass triples. 

We conclude that the rate of WD-WD collisions is likely to give rise
to only a small fraction of the type Ia SNe rate, of the order of
at most a few percents of the total type Ia SN rate in the more optimistic
case, and likely lower. Note, however, that the current data on triple
and higher multiplicity systems are still very limited, and more data
are needed in order provide better estimates, especially for collisions
of more massive WDs.

\section{Summary}

We investigated the result of head-on collisions of CO white dwarfs
with a helium layers as a channel for type Ia supernovae. Pure CO
WD-WD collisions were suggested in the past to be a channel for for
type Ia in globular clusters \citep{1989ApJ...342..986B,2009ApJ...705L.128R}
and more recently suggested as a channel for the production of type
Ia SNe from triple systems \citep{2011ApJ...741...82T,kat+12}. Simulations
of pure CO-WD collisions could have observational properties compatible
with observed type Ia SNe in terms of the observed Iron elements abundances
and energetics \citep{2013ApJ...778L..37K,2013MNRAS.436.3413G}. 

In this work we extend these studies to include WDs with a helium
layer on at least one of the WDs. A low mass helium shell of mass
up to $M_{{\rm He}}\le0.024M_{\odot}$ should exist in WDs according
to current stellar evolution models. More massive helium shells can
occur as a result of mass accretion from a companion. In these cases
a quadruple system is probably needed to lead to a collision (as He-rich
layer due to accretion requires a very close stellar companion). 

We studied WD-WD collisions using the FLASH code, exploring a range
of masses for both the WD CO bulk and for the outer Helium layers.
We find that collisions involving low mass He-layers $M_{{\rm He}}=0.01M_{\odot}$
give rise to He burning but no detonation occurs in the helium shell.
The helium shell partially burns into intermediate elements resulting
in a fast ejecta in the perpendicular plane to the collision. However,
the He-burning does not affect the overall evolution of the collision
and the CO-core detonation due to shock compression. In particular,
the nickel yields are very similar to those obtained from pure CO-WD
collisions. Nevertheless, high velocity material rich with $^{44}{\rm Ti}$,
and $^{48}{\rm Cr}$ are produced at an order of magnitude higher
levels than in pure CO-WD collision. We note that in general even
pure CO WD-WD collisions produce more of such intermediate elements
than single-degenerate models such as W7; the He-enriched collisions
therefore produce up to two orders of magnitude higher levels of such
intermediate elements compared with the single degenerate case.

Collisions with more massive He-layers ($M_{{\rm He}}=0.024M_{\odot}$)
due give rise to He-detonation, but the detonation front does not
propagate much before the CO-core detonates due to shock compression,
and the eventual outcomes of the collisions are still very similar
to those obtained for of $M_{{\rm He}}=0.01M_{\odot}$ cases. 

When the more He-rich collisions are considered ($M_{{\rm He}}=0.1M_{\odot}$; 
models 1F,2E)
the helium layer is sufficiently thick as to to enable the helium
detonation wave to propagate along the WD helium-shell. The helium
shell is burned mostly to intermediate elements and ejected with lower
velocities before a detonation is triggered in the CO core. The CO
detonation itself is still triggered by the shock compression and
behaves similarly to the case of pure CO-WD collisions, with almost
no effect of the helium burning itself on the CO detonation. The nickel
yields in this case are also similar to that of pure CO-WDs collisions.
The CO explosion ejects material with velocities higher than that
in ejected burned helium shell. This results with a second shock wave
accelerating the burned helium material to higher velocities.Finally,
in the most He-rich cases ($M_{{\rm He}}=0.2M_{\odot}$; model
1G), a converging
shock from the helium detonation is able to detonate the CO-core earlier,
before the shock compression due to the collision, resulting in a
lower production of Iron elements. 

We conclude that WD-WD collisions with He-layers may affect the outcome
of WD-WD collisions and give rise to significant composition changes,
as well as possible kinematic and morphological changes, which can
become significant for the most He-rich WD-WD collisions. We
note that after the submission of the current paper \cite{Kushnir+15}
studied these issues, finding similar outcomes and further confirming
our results.

\acknowledgements{}

HBP acknowledges support by the ISF I-CORE grant 1829/12 and the Technion
Deloro fellowship. The software used in this work was in part developed
by the DOE NNSA-ASC OASCR Flash Center at the University of Chicago.

\bibliographystyle{apj}
\bibliography{sample}

\begin{thebibliography}{50}
\expandafter\ifx\csname natexlab\endcsname\relax\def\natexlab#1{#1}\fi

\bibitem[{{Antonini} \& {Perets}(2012)}]{2012ApJ...757...27A}
{Antonini}, F. \& {Perets}, H.~B. 2012, \apj, 757, 27

\bibitem[{{Benz} {et~al.}(1989){Benz}, {Thielemann}, \&
  {Hills}}]{1989ApJ...342..986B}
{Benz}, W., {Thielemann}, F.-K., \& {Hills}, J.~G. 1989, \apj, 342, 986

\bibitem[{{Bildsten} {et~al.}(2007){Bildsten}, {Shen}, {Weinberg}, \&
  {Nelemans}}]{bil+07}
{Bildsten}, L., {Shen}, K.~J., {Weinberg}, N.~N., \& {Nelemans}, G. 2007,
  \apjl, 662, L95

\bibitem[{{Cyburt} {et~al.}(2010){Cyburt}, {Amthor}, {Ferguson}, {Meisel},
  {Smith}, {Warren}, {Heger}, {Hoffman}, {Rauscher}, {Sakharuk}, {Schatz},
  {Thielemann}, \& {Wiescher}}]{2010ApJS..189..240C}
{Cyburt}, R.~H., {Amthor}, A.~M., {Ferguson}, R., {Meisel}, Z., {Smith}, K.,
  {Warren}, S., {Heger}, A., {Hoffman}, R.~D., {Rauscher}, T., {Sakharuk}, A.,
  {Schatz}, H., {Thielemann}, F.~K., \& {Wiescher}, M. 2010, \apjs, 189, 240

\bibitem[{{Duch{\^e}ne} \& {Kraus}(2013)}]{duc+13}
{Duch{\^e}ne}, G. \& {Kraus}, A. 2013, \araa, 51, 269

\bibitem[{{Fekel}(1981)}]{fek81}
{Fekel}, Jr., F.~C. 1981, \apj, 246, 879

\bibitem[{{Fryxell} {et~al.}(2000){Fryxell}, {Olson}, {Ricker}, {Timmes},
  {Zingale}, {Lamb}, {MacNeice}, {Rosner}, {Truran}, \&
  {Tufo}}]{2000ApJS..131..273F}
{Fryxell}, B., {Olson}, K., {Ricker}, P., {Timmes}, F.~X., {Zingale}, M.,
  {Lamb}, D.~Q., {MacNeice}, P., {Rosner}, R., {Truran}, J.~W., \& {Tufo}, H.
  2000, \apjs, 131, 273

\bibitem[{{Garc{\'{\i}}a-Senz} {et~al.}(2013){Garc{\'{\i}}a-Senz},
  {Cabez{\'o}n}, {Arcones}, {Rela{\~n}o}, \&
  {Thielemann}}]{2013MNRAS.436.3413G}
{Garc{\'{\i}}a-Senz}, D., {Cabez{\'o}n}, R.~M., {Arcones}, A., {Rela{\~n}o},
  A., \& {Thielemann}, F.~K. 2013, \mnras, 436, 3413

\bibitem[{{Hamers} {et~al.}(2013){Hamers}, {Pols}, {Claeys}, \&
  {Nelemans}}]{2013MNRAS.430.2262H}
{Hamers}, A.~S., {Pols}, O.~R., {Claeys}, J.~S.~W., \& {Nelemans}, G. 2013,
  \mnras, 430, 2262

\bibitem[{{Hawley} {et~al.}(2012){Hawley}, {Athanassiadou}, \&
  {Timmes}}]{2012ApJ...759...39H}
{Hawley}, W.~P., {Athanassiadou}, T., \& {Timmes}, F.~X. 2012, \apj, 759, 39

\bibitem[{{Holcomb} \& {Kushnir}(2015)}]{Kushnir+15}
{Holcomb}, C. \& {Kushnir}, D. 2015, ArXiv e-prints

\bibitem[{{Jordan} {et~al.}(2012){Jordan}, {Perets}, {Fisher}, \& {van
  Rossum}}]{2012ApJ...761L..23J}
{Jordan}, IV, G.~C., {Perets}, H.~B., {Fisher}, R.~T., \& {van Rossum}, D.~R.
  2012, \apjl, 761, L23

\bibitem[{{Kashi} \& {Soker}(2011)}]{2011MNRAS.417.1466K}
{Kashi}, A. \& {Soker}, N. 2011, \mnras, 417, 1466

\bibitem[{{Katz} \& {Dong}(2012)}]{kat+12}
{Katz}, B. \& {Dong}, S. 2012, ArXiv e-prints

\bibitem[{{Kozai}(1962)}]{koz62}
{Kozai}, Y. 1962, \aj, 67, 591

\bibitem[{{Kromer} {et~al.}(2013){Kromer}, {Fink}, {Stanishev}, {Taubenberger},
  {Ciaraldi-Schoolman}, {Pakmor}, {R{\"o}pke}, {Ruiter}, {Seitenzahl}, {Sim},
  {Blanc}, {Elias-Rosa}, \& {Hillebrandt}}]{2013MNRAS.429.2287K}
{Kromer}, M., {Fink}, M., {Stanishev}, V., {Taubenberger}, S.,
  {Ciaraldi-Schoolman}, F., {Pakmor}, R., {R{\"o}pke}, F.~K., {Ruiter}, A.~J.,
  {Seitenzahl}, I.~R., {Sim}, S.~A., {Blanc}, G., {Elias-Rosa}, N., \&
  {Hillebrandt}, W. 2013, \mnras, 429, 2287

\bibitem[{{Krzemi{\'n}ski}(1972)}]{Krz+72}
{Krzemi{\'n}ski}, W. 1972, \emph{Acta Astron}, 22, 387

\bibitem[{{Kushnir} \& {Katz}(2014)}]{2014ApJ...785..124K}
{Kushnir}, D. \& {Katz}, B. 2014, \apj, 785, 124

\bibitem[{{Kushnir} {et~al.}(2013){Kushnir}, {Katz}, {Dong}, {Livne}, \&
  {Fern{\'a}ndez}}]{2013ApJ...778L..37K}
{Kushnir}, D., {Katz}, B., {Dong}, S., {Livne}, E., \& {Fern{\'a}ndez}, R.
  2013, \apjl, 778, L37

\bibitem[{{Lawlor} \& {MacDonald}(2006)}]{law+06}
{Lawlor}, T.~M. \& {MacDonald}, J. 2006, \mnras, 371, 263

\bibitem[{{Lidov}(1962)}]{lid62}
{Lidov}, M.~L. 1962, \planss, 9, 719

\bibitem[{{Livio} \& {Riess}(2003)}]{2003ApJ...594L..93L}
{Livio}, M. \& {Riess}, A.~G. 2003, \apjl, 594, L93

\bibitem[{{Livne} \& {Glasner}(1991)}]{liv+91}
{Livne}, E. \& {Glasner}, A.~S. 1991, \apj, 370, 272

\bibitem[{{Lopez} {et~al.}(2015){Lopez}, {Grefenstette}, {Reynolds}, {An},
  {Boggs}, {Christensen}, {Craig}, {Eriksen}, {Fryer}, {Hailey}, {Harrison},
  {Madsen}, {Stern}, {Zhang}, \& {Zoglauer}}]{Lop+15}
{Lopez}, L.~A., {Grefenstette}, B.~W., {Reynolds}, S.~P., {An}, H., {Boggs},
  S.~E., {Christensen}, F.~E., {Craig}, W.~W., {Eriksen}, K.~A., {Fryer},
  C.~L., {Hailey}, C.~J., {Harrison}, F.~A., {Madsen}, K.~K., {Stern}, D.~K.,
  {Zhang}, W.~W., \& {Zoglauer}, A. 2015, \apj, 814, 132

\bibitem[{{Lor{\'e}n-Aguilar} {et~al.}(2010){Lor{\'e}n-Aguilar}, {Isern}, \&
  {Garc{\'{\i}}a-Berro}}]{2010MNRAS.406.2749L}
{Lor{\'e}n-Aguilar}, P., {Isern}, J., \& {Garc{\'{\i}}a-Berro}, E. 2010,
  \mnras, 406, 2749

\bibitem[{{Maoz} {et~al.}(2014){Maoz}, {Mannucci}, \& {Nelemans}}]{mao+14}
{Maoz}, D., {Mannucci}, F., \& {Nelemans}, G. 2014, \araa, 52, 107

\bibitem[{{Michaely} \& {Perets}(2015)}]{mic+15}
{Michaely}, E. \& {Perets}, H.~B. 2015, ArXiv e-prints

\bibitem[{{Nomoto} {et~al.}(1984){Nomoto}, {Thielemann}, \&
  {Yokoi}}]{1984ApJ...286..644N}
{Nomoto}, K., {Thielemann}, F.-K., \& {Yokoi}, K. 1984, \apj, 286, 644

\bibitem[{{Pakmor} {et~al.}(2013){Pakmor}, {Kromer}, {Taubenberger}, \&
  {Springel}}]{2013ApJ...770L...8P}
{Pakmor}, R., {Kromer}, M., {Taubenberger}, S., \& {Springel}, V. 2013, \apjl,
  770, L8

\bibitem[{{Paxton} {et~al.}(2011){Paxton}, {Bildsten}, {Dotter}, {Herwig},
  {Lesaffre}, \& {Timmes}}]{2011ApJS..192....3P}
{Paxton}, B., {Bildsten}, L., {Dotter}, A., {Herwig}, F., {Lesaffre}, P., \&
  {Timmes}, F. 2011, \apjs, 192, 3

\bibitem[{{Paxton} {et~al.}(2015){Paxton}, {Marchant}, {Schwab}, {Bauer},
  {Bildsten}, {Cantiello}, {Dessart}, {Farmer}, {Hu}, {Langer}, {Townsend},
  {Townsley}, \& {Timmes}}]{2015ApJS..220...15P}
{Paxton}, B., {Marchant}, P., {Schwab}, J., {Bauer}, E.~B., {Bildsten}, L.,
  {Cantiello}, M., {Dessart}, L., {Farmer}, R., {Hu}, H., {Langer}, N.,
  {Townsend}, R.~H.~D., {Townsley}, D.~M., \& {Timmes}, F.~X. 2015, \apjs, 220,
  15

\bibitem[{{Perets}(2014)}]{per14}
{Perets}, H.~B. 2014, ArXiv e-prints

\bibitem[{{Perets} {et~al.}(2010){Perets}, {Gal-Yam}, {Mazzali}, {Arnett},
  {Kagan}, {Filippenko}, {Li}, {Arcavi}, {Cenko}, {Fox}, {Leonard}, {Moon},
  {Sand}, {Soderberg}, {Anderson}, {James}, {Foley}, {Ganeshalingam}, {Ofek},
  {Bildsten}, {Nelemans}, {Shen}, {Weinberg}, {Metzger}, {Piro}, {Quataert},
  {Kiewe}, \& {Poznanski}}]{per+10}
{Perets}, H.~B., {Gal-Yam}, A., {Mazzali}, P.~A., {Arnett}, D., {Kagan}, D.,
  {Filippenko}, A.~V., {Li}, W., {Arcavi}, I., {Cenko}, S.~B., {Fox}, D.~B.,
  {Leonard}, D.~C., {Moon}, D.-S., {Sand}, D.~J., {Soderberg}, A.~M.,
  {Anderson}, J.~P., {James}, P.~A., {Foley}, R.~J., {Ganeshalingam}, M.,
  {Ofek}, E.~O., {Bildsten}, L., {Nelemans}, G., {Shen}, K.~J., {Weinberg},
  N.~N., {Metzger}, B.~D., {Piro}, A.~L., {Quataert}, E., {Kiewe}, M., \&
  {Poznanski}, D. 2010, \nat, 465, 322

\bibitem[{{Raskin} {et~al.}(2012){Raskin}, {Scannapieco}, {Fryer},
  {Rockefeller}, \& {Timmes}}]{2012ApJ...746...62R}
{Raskin}, C., {Scannapieco}, E., {Fryer}, C., {Rockefeller}, G., \& {Timmes},
  F.~X. 2012, \apj, 746, 62

\bibitem[{{Raskin} {et~al.}(2010){Raskin}, {Scannapieco}, {Rockefeller},
  {Fryer}, {Diehl}, \& {Timmes}}]{2010ApJ...724..111R}
{Raskin}, C., {Scannapieco}, E., {Rockefeller}, G., {Fryer}, C., {Diehl}, S.,
  \& {Timmes}, F.~X. 2010, \apj, 724, 111

\bibitem[{{Raskin} {et~al.}(2009){Raskin}, {Timmes}, {Scannapieco}, {Diehl}, \&
  {Fryer}}]{2009MNRAS.399L.156R}
{Raskin}, C., {Timmes}, F.~X., {Scannapieco}, E., {Diehl}, S., \& {Fryer}, C.
  2009, \mnras, 399, L156

\bibitem[{{Riddle} {et~al.}(2015){Riddle}, {Tokovinin}, {Mason}, {Hartkopf},
  {Roberts}, {Baranec}, {Law}, {Bui}, {Burse}, {Das}, {Dekany}, {Kulkarni},
  {Punnadi}, {Ramaprakash}, \& {Tendulkar}}]{rid+15}
{Riddle}, R.~L., {Tokovinin}, A., {Mason}, B.~D., {Hartkopf}, W.~I., {Roberts},
  Jr., L.~C., {Baranec}, C., {Law}, N.~M., {Bui}, K., {Burse}, M.~P., {Das},
  H.~K., {Dekany}, R.~G., {Kulkarni}, S., {Punnadi}, S., {Ramaprakash}, A.~N.,
  \& {Tendulkar}, S.~P. 2015, \apj, 799, 4

\bibitem[{{Rosswog} {et~al.}(2009){Rosswog}, {Kasen}, {Guillochon}, \&
  {Ramirez-Ruiz}}]{2009ApJ...705L.128R}
{Rosswog}, S., {Kasen}, D., {Guillochon}, J., \& {Ramirez-Ruiz}, E. 2009,
  \apjl, 705, L128

\bibitem[{{Sana} {et~al.}(2013){Sana}, {de Koter}, {de Mink}, {Dunstall},
  {Evans}, {H{\'e}nault-Brunet}, {Ma{\'{\i}}z Apell{\'a}niz},
  {Ram{\'{\i}}rez-Agudelo}, {Taylor}, {Walborn}, {Clark}, {Crowther},
  {Herrero}, {Gieles}, {Langer}, {Lennon}, \& {Vink}}]{san+13}
{Sana}, H., {de Koter}, A., {de Mink}, S.~E., {Dunstall}, P.~R., {Evans},
  C.~J., {H{\'e}nault-Brunet}, V., {Ma{\'{\i}}z Apell{\'a}niz}, J.,
  {Ram{\'{\i}}rez-Agudelo}, O.~H., {Taylor}, W.~D., {Walborn}, N.~R., {Clark},
  J.~S., {Crowther}, P.~A., {Herrero}, A., {Gieles}, M., {Langer}, N.,
  {Lennon}, D.~J., \& {Vink}, J.~S. 2013, \aap, 550, A107

\bibitem[{{Shen} \& {Bildsten}(2009)}]{she+09}
{Shen}, K.~J. \& {Bildsten}, L. 2009, \apj, 699, 1365

\bibitem[{{Shen} \& {Bildsten}(2014)}]{2014ApJ...785...61S}
---. 2014, \apj, 785, 61

\bibitem[{{Soker} {et~al.}(2014){Soker}, {Garc{\'{\i}}a-Berro}, \&
  {Althaus}}]{sok+14}
{Soker}, N., {Garc{\'{\i}}a-Berro}, E., \& {Althaus}, L.~G. 2014, \mnras, 437,
  L66

\bibitem[{{Thompson}(2011)}]{2011ApJ...741...82T}
{Thompson}, T.~A. 2011, \apj, 741, 82

\bibitem[{{Timmes} {et~al.}(2000){Timmes}, {Hoffman}, \&
  {Woosley}}]{2000ApJS..129..377T}
{Timmes}, F.~X., {Hoffman}, R.~D., \& {Woosley}, S.~E. 2000, \apjs, 129, 377

\bibitem[{{Timmes} \& {Swesty}(2000)}]{2000ApJS..126..501T}
{Timmes}, F.~X. \& {Swesty}, F.~D. 2000, \apjs, 126, 501

\bibitem[{{Tokovinin}(2014)}]{tok14}
{Tokovinin}, A. 2014, \aj, 147, 87

\bibitem[{{Troja} {et~al.}(2014){Troja}, {Segreto}, {La Parola}, {Hartmann},
  {Baumgartner}, {Markwardt}, {Barthelmy}, {Cusumano}, \& {Gehrels}}]{tro+14}
{Troja}, E., {Segreto}, A., {La Parola}, V., {Hartmann}, D., {Baumgartner}, W.,
  {Markwardt}, C., {Barthelmy}, S., {Cusumano}, G., \& {Gehrels}, N. 2014,
  \apjl, 797, L6

\bibitem[{{Waldman} {et~al.}(2011){Waldman}, {Sauer}, {Livne}, {Perets},
  {Glasner}, {Mazzali}, {Truran}, \& {Gal-Yam}}]{wal+11}
{Waldman}, R., {Sauer}, D., {Livne}, E., {Perets}, H., {Glasner}, A.,
  {Mazzali}, P., {Truran}, J.~W., \& {Gal-Yam}, A. 2011, \apj, 738, 21

\bibitem[{{Wolf} {et~al.}(2013){Wolf}, {Bildsten}, {Brooks}, \&
  {Paxton}}]{2013ApJ...777..136W}
{Wolf}, W.~M., {Bildsten}, L., {Brooks}, J., \& {Paxton}, B. 2013, \apj, 777,
  136

\bibitem[{{Woosley} {et~al.}(1986){Woosley}, {Taam}, \& {Weaver}}]{woo+86}
{Woosley}, S.~E., {Taam}, R.~E., \& {Weaver}, T.~A. 1986, \apj, 301, 601

\end{thebibliography}

\newpage

\begin{longtable}{cccccccccc}
\hline
\hline

\\
\\
\vspace{2pt} 
\\

Isotope  & 1A &   & 1C & & Isotope &  1A &   & 1C \\

\hline
\\

$^{1}$H & 2.23e-04 & 2.25e-04 & 2.37e-04 & 2.36e-04 & $^{46}$Ti &  4.63e-08 &  3.37e-08 &  7.12e-08 &  7.12e-08 \\
Neut &  2.95e-16 &  2.30e-35 &  1.22e-22 &  3.17e-34   & $^{47}$Ti &  4.19e-10 &  2.18e-07 &  4.71e-10 &  6.86e-07 \\
$^{2}$H &  3.55e-16 &  6.29e-17 &  3.10e-15 &  1.56e-15  &  $^{48}$Ti &  3.59e-09 &  5.68e-04 &  3.63e-12 &  8.17e-04 \\
$^{3}$He &  1.71e-13 &  1.89e-13 &  1.22e-12 &  1.21e-12  &  $^{49}$Ti &  2.15e-13 &  1.27e-05 &  4.96e-15 &  1.62e-05 \\
$^{4}$He &  5.47e-03 &  5.55e-03 &  8.46e-03 &  8.46e-03  &  $^{50}$Ti &  6.15e-18 &  1.03e-17 &  1.26e-17 &  1.26e-17 \\
$^{7}$Li &  1.72e-15 &  5.93e-16 &  7.36e-17 &  9.90e-16  &  $^{51}$Ti &  3.03e-24 &  1.29e-33 &  7.56e-24 &  1.89e-32 \\
$^{7}$Be &  2.01e-10 &  2.01e-10 &  1.60e-10 &  1.59e-10  &  $^{52}$Ti &  3.69e-27 &  1.95e-34 &  5.12e-27 &  3.14e-33 \\
$^{9}$Be &  4.35e-15 &  2.28e-15 &  1.57e-14 &  1.57e-14  &  $^{53}$Ti &  6.37e-29 &  6.39e-35 &  6.40e-29 &  1.07e-33 \\
$^{10}$Be &  9.10e-22 &  1.55e-23 &  1.34e-21 &  3.56e-25  & $^{54}$Ti &  6.49e-29 &  2.99e-36 &  6.49e-29 &  5.08e-35 \\
$^{8}$B &  3.15e-13 &  4.12e-19 &  1.17e-14 &  2.88e-22  &   $^{47}$V &  2.14e-07 &  6.59e-33 &  6.60e-07 &  9.54e-32 \\
$^{12}$C &  5.04e-02 &  9.25e-02 &  2.76e-02 &  2.76e-02  &  $^{48}$V &  7.97e-09 &  3.02e-28 &  7.49e-09 &  9.43e-29 \\
$^{13}$C &  3.54e-10 &  3.89e-08 &  3.36e-10 &  6.17e-08  &  $^{49}$V &  2.30e-08 &  2.30e-20 &  9.81e-09 &  4.41e-19 \\
$^{13}$N &  4.57e-08 &  2.84e-34 &  6.01e-08 &  3.93e-33  &  $^{50}$V &  2.29e-14 &  1.73e-14 &  2.57e-14 &  2.57e-14 \\
$^{14}$N &  1.12e-07 &  1.49e-06 &  1.16e-07 &  1.97e-06  &  $^{51}$V &  9.53e-14 &  7.69e-08 &  2.82e-14 &  3.70e-07 \\
$^{15}$N &  4.58e-09 &  1.03e-07 &  3.97e-09 &  3.30e-07  &  $^{52}$V &  3.11e-19 &  8.59e-34 &  4.96e-19 &  1.28e-32 \\
$^{14}$O &  1.67e-06 &  3.64e-35 &  1.86e-06 &  5.95e-34  &  $^{53}$V &  7.24e-24 &  3.61e-34 &  2.54e-23 &  5.78e-33 \\
$^{15}$O &  1.33e-07 &  6.74e-35 &  2.47e-07 &  1.07e-33  &  $^{54}$V &  2.99e-28 &  1.98e-34 &  3.24e-28 &  3.27e-33 \\
$^{16}$O &  1.01e-01 &  1.26e-01 &  9.66e-02 &  9.66e-02  &  $^{55}$V &  4.44e-26 &  1.33e-35 &  5.58e-26 &  2.25e-34 \\
$^{17}$O &  4.66e-09 &  1.18e-07 &  1.64e-09 &  2.16e-07  &  $^{56}$V &  6.73e-29 &  4.46e-37 &  6.85e-29 &  7.59e-36 \\
$^{18}$O &  1.69e-12 &  8.95e-10 &  7.35e-14 &  9.02e-10  &  $^{47}$Cr &  7.07e-11 &  8.67e-37 &  2.56e-08 &  1.48e-35 \\
$^{17}$F &  1.20e-07 &  4.04e-35 &  9.84e-08 &  6.62e-34  &  $^{48}$Cr &  5.91e-04 &  1.05e-31 &  8.17e-04 &  1.47e-30 \\
$^{18}$F &  5.41e-09 &  8.97e-33 &  2.77e-10 &  1.16e-31  &  $^{49}$Cr &  1.36e-05 &  4.42e-33 &  1.62e-05 &  5.60e-32 \\
$^{19}$F &  2.63e-10 &  1.95e-09 &  4.23e-10 &  4.45e-09  &  $^{50}$Cr &  4.37e-07 &  3.01e-07 &  3.00e-07 &  3.00e-07 \\
$^{18}$Ne &  2.79e-09 &  1.11e-36 &  1.71e-08 &  1.89e-35  & $^{51}$Cr &  5.00e-10 &  1.04e-29 &  5.18e-10 &  3.88e-28 \\
$^{19}$Ne &  1.80e-09 &  1.21e-35 &  4.95e-09 &  2.03e-34  & $^{52}$Cr &  7.46e-09 &  9.35e-03 &  3.01e-11 &  1.12e-02 \\
$^{20}$Ne &  1.04e-03 &  9.04e-04 &  2.02e-03 &  2.02e-03  & $^{53}$Cr &  2.31e-12 &  3.17e-04 &  6.77e-15 &  2.37e-08 \\
$^{21}$Ne &  3.32e-09 &  3.03e-09 &  1.07e-08 &  1.07e-08  & $^{54}$Cr &  2.84e-17 &  2.12e-13 &  5.41e-17 &  2.71e-13 \\
$^{22}$Ne &  3.35e-13 &  3.10e-13 &  2.38e-12 &  2.38e-12  & $^{55}$Cr &  7.83e-24 &  8.49e-34 &  1.79e-23 &  1.30e-32 \\
$^{23}$Na &  1.25e-06 &  2.74e-06 &  1.38e-06 &  4.87e-06  & $^{56}$Cr &  1.33e-26 &  1.46e-33 &  4.82e-25 &  2.13e-32 \\
$^{24}$Na &  1.01e-13 &  3.75e-32 &  9.98e-13 &  1.84e-29  & $^{57}$Cr &  7.37e-26 &  8.05e-26 &  2.82e-24 &  2.82e-24 \\
$^{23}$Mg &  1.94e-06 &  9.59e-36 &  3.50e-06 &  1.62e-34  & $^{58}$Cr &  6.96e-29 &  2.52e-26 &  1.92e-20 &  1.92e-20 \\
$^{24}$Mg &  2.87e-03 &  2.15e-03 &  5.45e-03 &  5.45e-03  & $^{51}$Mn &  9.00e-08 &  1.00e-32 &  3.69e-07 &  1.45e-31 \\
$^{25}$Mg &  1.10e-06 &  9.41e-07 &  1.90e-06 &  1.90e-06  & $^{52}$Mn &  9.26e-07 &  1.31e-30 &  4.32e-07 &  1.78e-29 \\
$^{27}$Al &  4.59e-06 &  4.04e-06 &  6.96e-06 &  8.99e-06  & $^{53}$Mn &  2.87e-06 &  6.65e-16 &  1.23e-06 &  4.04e-04 \\
$^{27}$Si &  4.77e-07 &  4.13e-36 &  2.03e-06 &  7.02e-35  & $^{54}$Mn &  2.91e-13 &  7.41e-23 &  2.71e-13 &  1.03e-22 \\
$^{28}$Si &  3.10e-01 &  2.79e-01 &  3.12e-01 &  3.12e-01  & $^{55}$Mn &  2.17e-13 &  1.87e-06 &  5.92e-14 &  1.32e-06 \\
$^{30}$P &  6.86e-05 &  4.31e-05 &  8.03e-05 &  8.03e-05  &  $^{56}$Mn &  5.27e-19 &  5.29e-32 &  6.69e-19 &  7.27e-31 \\
$^{31}$P &  2.35e-05 &  2.23e-05 &  1.23e-05 &  2.45e-05  &  $^{51}$Fe &  2.98e-15 &  5.74e-37 &  5.19e-11 &  9.77e-36 \\
$^{31}$S &  4.76e-06 &  2.94e-36 &  1.28e-05 &  4.99e-35  &  $^{52}$Fe &  9.79e-03 &  4.68e-32 &  1.12e-02 &  6.36e-31 \\
$^{32}$S &  1.94e-01 &  1.78e-01 &  1.91e-01 &  1.91e-01  &  $^{53}$Fe &  3.30e-04 &  9.91e-34 &  4.03e-04 &  1.39e-32 \\
$^{35}$Cl &  2.69e-06 &  2.53e-06 &  3.33e-06 &  5.00e-06  & $^{54}$Fe &  7.49e-06 &  6.12e-06 &  2.91e-06 &  2.91e-06 \\
$^{36}$Cl &  9.00e-10 &  3.08e-23 &  1.38e-09 &  1.38e-09  & $^{55}$Fe &  7.98e-09 &  4.12e-28 &  4.95e-09 &  9.92e-19 \\
$^{37}$Cl &  1.27e-09 &  2.96e-06 &  1.77e-09 &  5.36e-06  & $^{56}$Fe &  3.62e-05 &  4.21e-01 &  5.29e-10 &  4.51e-01 \\
$^{38}$Cl &  6.05e-17 &  3.02e-33 &  8.16e-17 &  3.84e-32  & $^{57}$Fe &  9.12e-07 &  4.55e-03 &  4.23e-14 &  4.86e-03 \\
$^{35}$Ar &  2.23e-07 &  2.30e-36 &  1.78e-06 &  3.91e-35  & $^{58}$Fe &  1.47e-15 &  1.81e-13 &  1.94e-15 &  2.76e-13 \\
$^{36}$Ar &  4.14e-02 &  3.82e-02 &  4.13e-02 &  4.13e-02  & $^{59}$Fe &  2.73e-21 &  4.88e-30 &  1.13e-19 &  4.84e-29 \\
$^{37}$Ar &  4.47e-06 &  2.99e-27 &  5.36e-06 &  1.12e-20  & $^{60}$Fe &  1.40e-18 &  1.04e-28 &  4.68e-16 &  4.68e-16 \\
$^{38}$Ar &  5.26e-07 &  3.55e-07 &  6.84e-07 &  6.84e-07  & $^{61}$Fe &  6.42e-21 &  8.03e-34 &  6.61e-21 &  1.17e-32 \\
$^{39}$Ar &  7.42e-13 &  3.29e-26 &  1.20e-12 &  5.29e-13  & $^{62}$Fe &  3.02e-16 &  1.55e-34 &  3.89e-16 &  2.54e-33 \\
$^{40}$Ar &  6.97e-15 &  1.70e-12 &  1.28e-14 &  1.28e-14  & $^{63}$Fe &  1.87e-21 &  1.42e-35 &  4.61e-18 &  2.40e-34 \\
$^{41}$Ar &  2.72e-21 &  9.15e-33 &  4.70e-21 &  1.17e-31  & $^{64}$Fe &  4.14e-17 &  4.72e-36 &  2.62e-14 &  8.03e-35 \\
$^{39}$K &  1.79e-06 &  1.33e-06 &  2.76e-06 &  2.94e-06  &  $^{65}$Fe &  1.56e-23 &  1.94e-36 &  3.67e-19 &  3.30e-35 \\
$^{40}$K &  3.02e-11 &  3.25e-12 &  4.02e-11 &  4.02e-11  &  $^{66}$Fe &  7.95e-29 &  1.07e-36 &  2.52e-24 &  1.82e-35 \\
$^{41}$K &  3.01e-12 &  5.61e-07 &  3.66e-12 &  1.83e-09  &  $^{55}$Co &  1.53e-06 &  2.15e-31 &  1.32e-06 &  2.92e-30 \\
$^{42}$K &  4.00e-17 &  5.48e-32 &  5.87e-17 &  7.54e-31  &  $^{56}$Co &  9.30e-07 &  1.58e-29 &  7.99e-07 &  1.15e-25 \\
$^{43}$K &  5.11e-21 &  9.72e-32 &  1.40e-20 &  1.36e-30  &  $^{57}$Co &  7.98e-08 &  4.85e-20 &  3.48e-08 &  4.85e-20 \\
$^{44}$K &  3.79e-26 &  2.11e-33 &  1.51e-25 &  2.74e-32  &  $^{58}$Co &  2.93e-13 &  2.34e-23 &  2.77e-13 &  9.86e-29 \\
$^{39}$Ca &  3.16e-09 &  1.24e-36 &  1.90e-07 &  2.11e-35  & $^{59}$Co &  8.84e-11 &  2.46e-07 &  3.49e-13 &  1.00e-09 \\
$^{40}$Ca &  4.04e-02 &  3.75e-02 &  4.14e-02 &  4.14e-02  & $^{60}$Co &  1.09e-17 &  7.92e-29 &  1.53e-16 &  1.03e-21 \\
$^{41}$Ca &  7.73e-07 &  3.36e-20 &  8.51e-07 &  8.50e-07  & $^{61}$Co &  1.63e-17 &  2.47e-32 &  8.71e-15 &  3.36e-31 \\
$^{42}$Ca &  1.41e-08 &  9.59e-09 &  2.32e-08 &  2.32e-08  & $^{62}$Co &  2.89e-17 &  4.11e-34 &  1.58e-17 &  6.52e-33 \\
$^{43}$Ca &  1.23e-10 &  2.57e-07 &  4.54e-10 &  4.55e-06  & $^{63}$Co &  2.15e-16 &  1.27e-34 &  1.28e-14 &  2.13e-33 \\
$^{44}$Ca &  2.44e-09 &  4.92e-05 &  6.52e-13 &  8.86e-05  & $^{64}$Co &  9.01e-18 &  1.42e-36 &  6.08e-15 &  2.41e-35 \\
$^{45}$Ca &  3.37e-17 &  1.55e-28 &  7.09e-17 &  1.25e-28  & $^{65}$Co &  2.34e-21 &  5.56e-36 &  1.53e-17 &  9.46e-35 \\
$^{46}$Ca &  5.38e-20 &  2.85e-16 &  1.53e-19 &  8.73e-16  & $^{66}$Co &  7.94e-29 &  9.74e-37 &  2.10e-24 &  1.66e-35 \\
$^{47}$Ca &  3.35e-25 &  4.64e-31 &  1.74e-24 &  6.56e-30  & $^{67}$Co &  8.06e-29 &  1.05e-36 &  8.15e-29 &  1.79e-35 \\
$^{48}$Ca &  3.96e-27 &  1.73e-26 &  6.87e-27 &  6.87e-27  & $^{55}$Ni &  9.23e-19 &  4.16e-37 &  3.03e-12 &  7.08e-36 \\
$^{49}$Ca &  5.94e-29 &  9.37e-34 &  5.98e-29 &  1.31e-32  & $^{56}$Ni &  4.33e-01 &  7.27e-31 &  4.51e-01 &  1.01e-29 \\
$^{43}$Sc &  2.53e-07 &  3.92e-32 &  4.48e-06 &  5.55e-31  & $^{57}$Ni &  4.55e-03 &  1.99e-31 &  4.86e-03 &  2.82e-30 \\
$^{44}$Sc &  1.17e-11 &  2.06e-32 &  1.70e-11 &  1.76e-11  & $^{58}$Ni &  4.58e-03 &  4.62e-03 &  4.79e-03 &  4.79e-03 \\
$^{45}$Sc &  3.95e-11 &  5.63e-08 &  4.64e-11 &  1.41e-07  & $^{59}$Ni &  2.43e-07 &  1.62e-18 &  3.48e-07 &  3.47e-07 \\
$^{46}$Sc &  8.90e-16 &  3.66e-30 &  1.75e-15 &  8.12e-29  & $^{60}$Ni &  3.41e-09 &  3.31e-09 &  8.94e-09 &  8.94e-09 \\
$^{47}$Sc &  8.91e-18 &  6.71e-31 &  1.62e-17 &  9.43e-30  & $^{61}$Ni &  9.04e-14 &  5.07e-14 &  3.36e-13 &  3.45e-13 \\
$^{48}$Sc &  4.12e-22 &  2.02e-31 &  1.66e-21 &  2.88e-30  & $^{62}$Ni &  3.18e-14 &  2.17e-14 &  2.67e-13 &  2.68e-13 \\
$^{49}$Sc &  4.04e-24 &  1.18e-32 &  1.07e-23 &  1.56e-31  & $^{63}$Ni &  5.75e-17 &  2.72e-16 &  1.69e-15 &  1.45e-14 \\
$^{50}$Sc &  6.01e-29 &  1.88e-34 &  6.03e-29 &  3.02e-33  & $^{64}$Ni &  1.09e-15 &  1.15e-15 &  9.26e-14 &  1.25e-13 \\
$^{51}$Sc &  6.12e-29 &  2.33e-35 &  6.12e-29 &  3.94e-34  & $^{65}$Ni &  1.17e-19 &  1.11e-19 &  7.60e-17 &  9.16e-17 \\
$^{43}$Ti &  3.72e-11 &  8.07e-37 &  7.36e-08 &  1.37e-35  & $^{66}$Ni &  4.87e-21 &  3.53e-20 &  3.39e-17 &  3.39e-17 \\
$^{44}$Ti &  5.10e-05 &  5.32e-29 &  9.09e-05 &  2.32e-06  & $^{67}$Ni &  1.26e-24 &  3.00e-24 &  1.57e-23 &  1.70e-23 \\
$^{45}$Ti &  6.82e-08 &  1.63e-32 &  1.41e-07 &  2.12e-31  & $^{68}$Ni &  2.12e-27 &  2.12e-27 &  5.63e-20 &  5.63e-20 \\
\hline
\caption{Post-processing nucleosynthetic results for models 1A and 1C. The first column for each model is at the end of the simulation. The second column for each model are the abundances after $10 ~\rm Gy$. All results are in solar mass.}
\label{post-processing}
\end{longtable}
\end{document}